\documentclass[10.5pt,a4paper,aps,prb,preprintnumbers,amsmath,amssymb,twocolumn,longbibliography,superscriptaddress]{revtex4-2}

\usepackage{amsmath}
\usepackage{amssymb}
\usepackage{mathtools}
\usepackage[utf8]{inputenc}
\usepackage[T1]{fontenc}
\usepackage{graphicx}
\usepackage{braket}
\usepackage{physics}
\usepackage{here}
\usepackage{wrapfig}
\usepackage{color}
\usepackage{array}
\usepackage{comment}

\begin{document}
	\title{Contemporary tensor network approaches to gapless and topological phases in an extended Bose-Hubbard ladder}
	
	\author{Yuma Watanabe}
	\affiliation{ICFO-Institut de Ciencies Fotoniques, The Barcelona Institute of Science and Technology, Av. Carl Friedrich Gauss 3, 08860 Castelldefels (Barcelona), Spain}
	\author{Ravindra W.  Chhajlany}
	\affiliation{Institute of Spintronics and Quantum Information, Faculty of Physics and Astronomy, Adam Mickiewicz University, 61614 Poznan, Poland}
	\author{Maciej Lewenstein}
	\affiliation{ICFO-Institut de Ciencies Fotoniques, The Barcelona Institute of Science and Technology, 08860 Castelldefels (Barcelona), Spain}
	\affiliation{ICREA, Pg. Lluis Companys 23, 08010 Barcelona, Spain}
	\author{Tobias Gra\ss}
	\affiliation{DIPC - Donostia International Physics Center, Paseo Manuel de Lardiz{\'a}bal 4, 20018 San Sebasti{\'a}n, Spain}
	\affiliation{IKERBASQUE, Basque Foundation for Science, Plaza Euskadi 5, 48009 Bilbao, Spain}
	\author{Utso Bhattacharya}
	\affiliation{Institute for Theoretical Physics, ETH Zurich, Zurich, Switzerland}
	\begin{abstract}
		The development of numerically efficient computational methods has facilitated in depth studies of various correlated phases of matter including critical and topological phases. A quantum Monte-Carlo study of an extended Bose-Hubbard ladder has recently been used to identify an exotic phase with hidden order, where superfluid correlations coexist with string order, dubbed a Haldane superfluid (HSF). However, finite-size methods can struggle to uniquely determine the boundaries of quasi-long-range ordered states with nonlocal, e.g. string-like, correlations. In the present Letter, we revisit the HSF scenario using  tensor network algorithms specialized for finite/infinite (quasi-)1D systems, \textit{i.e.} the well-governed finite-size density matrix renormalization group (DMRG), and the state-of-the-art infinite-size variational uniform matrix product state (VUMPS) methods. While DMRG results extrapolated to the thermodynamic limit are compatible with a putative HSF, the results from the VUMPS calculations provide sharper phase boundaries that leave no room for such a topological superfluid. Our results demonstrate the crucial advantage of the VUMPS in characterizing topological and critical interacting phases providing the precise phase boundaries.
	\end{abstract}
	\maketitle
	
	{\it Introduction. } 
The study of novel exotic equilibrium phases is one of the frontiers in many-body physics.
The most challenging task is to evade the exponential growth of Hilbert space, while precisely and sufficiently capturing the ground/low-excited state properties.
To this end, various numerical methods have been proposed, such as exact diagonalization,  quantum Monte Carlo (QMC)~\cite{fehske_computational_2008}, tensor network algorithms \cite{schollwock_density-matrix_2011} and machine learning inspired approaches ~\cite{dawid_modern_2023}.
In particular, the tensor network (TN) approach provides access to large system sizes without model-dependent limitations, while capturing relevant correlations at a reasonable computational cost.
The celebrated density matrix renormalization group (DMRG)~\cite{white_density_1992} is a leading example of this class of methods wherein low energy states are obtained effectively as variationally optimised matrix product states (MPS). 
Extended TN methods specialized for specific motivations have been proposed, such as time-evolving block decimation (TEBD)\cite{vidal_efficient_2003,vidal_efficient_2004} for (real/imaginary) time evolution, projected entangled pair states (PEPS)\cite{verstraete_renormalization_2004,orus_practical_2014} for 2D systems, as well as the multiscale entanglement renormalization ansatz (MERA)~\cite{vidal_entanglement_2007,vidal_class_2008} for critical systems.
The TN method also provides access to infinite systems where the size/boundary effects can be discarded.
The state-of-the-art infinite TN is the variational uniform matrix product state (VUMPS)~\cite{zauner-stauber_variational_2018}, which eliminates the disadvantages of conventional methods, such as the infinite DMRG/TEBD (iDMRG/iTEBD)~\cite{vidal_efficient_2003,vidal_entanglement_2007,mcculloch_infinite_2008}; namely it directly deals with infinite systems by variationally optimizing uniform MPS.

Interesting target phases for TNs are topological phases~\cite {senthil_symmetry-protected_2015,wen_colloquium_2017} in interacting 1D/quasi-1D/2D systems.
The simplest example of a topological phase is the so-called symmetry-protected-topological (SPT) phase~\cite{senthil_symmetry-protected_2015} 
Despite significant advances in the study of non-interacting SPT~\cite{schnyder_classification_2008,kitaev_periodic_2009,bernevig_topological_2013}, a comprehensive understanding of interacting/interaction-induced SPT~\cite{rachel_interacting_2018} is still challenging, remaining a hot topic in modern physics.
TN approaches facilitate access to crucial quantities characterizing interacting SPT phases, such as the non-local string order~\cite{oshikawa_hidden_1992},  degenerate edge modes and the degeneracy in the entanglement spectrum~\cite{li_entanglement_2008,pollmann_entanglement_2010,pollmann_symmetry_2012}, which can be difficult to access with other methods, such as the QMC.

Interacting SPT phases are hosted by the 1D Bose-Hubbard (BH) model with non-local interactions~\cite{goral_quantum_2002,sengupta_supersolids_2005,batrouni_supersolid_2006,dalla_torre_hidden_2006,mishra_supersolid_2009,iskin_route_2011,kimura_gutzwiller_2011,rossini_phase_2012,ohgoe_ground-state_2012,bogner_variational_2019}. 
This model has a richer phase diagram than the standard BH model~\cite{jaksch_cold_1998} where the competition between hopping of bosonic particles and on-site repulsion leads to a quantum phase transition only between a superfluid (SF) phase and a Mott-insulating (MI) phase, observed experimentally with cold atoms in optical lattices ~\cite{greiner_quantum_2002}. 
In the extended BH model~\cite{dutta_non-standard_2015}, recently realized with dipolar excitons on a synthetic square lattice ~\cite{lagoin_extended_2022} and with dipolar atoms~\cite{su_dipolar_2023}, other insulating phases appear, including insulators with density wave (DW) modulations, and in 1D an insulating SPT phase, known as a Haldane insulator (HI)~\cite{dalla_torre_hidden_2006} characterized by nonlocal correlations. 
In contrast to the DW insulator, the HI phase respects the discrete translational symmetry of the lattice. 
Nevertheless, a peculiar DW-like pattern of alternating high- and low-density sites is present, hidden by strings of arbitrary numbers of sites with average density. 
This hidden string ordered phase is the bosonic analog of the spin-1 Haldane phase and is an interacting SPT phase similarly characterized by localized gapless edge excitations and degeneracy in the entanglement spectrum~\cite{ejima_spectral_2014}.

An interesting question concerns the possible coexistence of features of these insulators, such as DW or string order, and superfluid correlations. 
In the case of DW order, the coexistence with superfluid correlations gives rise to an exotic phase known as the supersolid phase, which has recently been observed with cold atoms~\cite{guo_low-energy_2019,norcia_two-dimensional_2021,chisholm_probing_2024}. 
The coexistence of superfluidity and Haldane string order has only been reported theoretically in Ref.~\onlinecite{lv_exotic_2016}, where Monte-Carlo simulation suggests that such a Haldane superfluid (HSF) arises by coupling two BH chains with nearest-neighbor interactions in a parameter regime where the independent chains would be in the SF phase. 
Such two-leg ladders have recently been studied also from the point of view of supersolidity  at arbitrary filling factors~\cite{watanabe_competing_2024}, and novel topological phases~\cite{wellnitz_emergent_2024,korbmacher_topological_2025}. Experimentally, a cold atom Fermi-Hubbard two-leg ladder has allowed for realizing the string order in the Haldane phase~\cite{sompet_realizing_2022}.

However, from both experimental and theoretical points of view, the precise determination of the phase diagram is extremely challenging, as the boundaries of string order and quasi-long-range SF correlations in 1D systems are washed out through finite-size effects. 
It raises a question regarding the existence of the HSF phase reported by the QMC simulation~\cite{lv_exotic_2016}, which is supported only by the scanty string order.
Therefore, this Letter revisits the scenario using TN methods for (quasi-)1D finite/infinite systems.
Specifically, we use the DMRG and the VUMPS methods for finite/infinite systems using the ITensor library~\cite{fishman_itensor_2022} and primarily focus on obtaining the precise transition points by studying non-local correlations.
We show that only the VUMPS calculations provide very sharp boundaries, which are crucial to accurately determine the different phases of the model.
While the DMRG results exhibit regimes with overlapping string and SF order, providing room for the putative Haldane superfluid (pHSF) phase, the string order in the VUMPS calculation sharply drops to zero in the parameter regime where superfluid correlations appear in the system. 
This finding showcases the crucial advantage of this infinite-system tensor network method in characterizing topological and critical phases.
The investigations of the topological properties regarding the edge mode and the entanglement spectrum (provided in the supplemental material) further support the VUMPS results.

	{\it Model and Methods.}
	\begin{figure}
		\centering
		\includegraphics[scale=0.6]{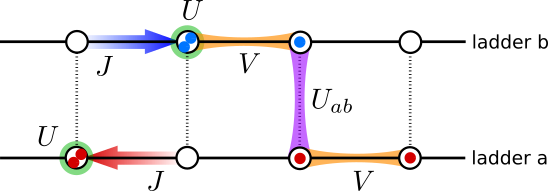}
		\caption{
			The schematic of the BH model in a two-leg ladder.
			Each leg has the nearest-neighbor hopping $J$, the on-site interaction $U$, and the nearest-neighbor interactions $V$.
			Particles on different legs but on the same rung interact via  $U_{ab}$.
		}
		\label{fig:model}
	\end{figure}
	Here, we consider the one-dimensional two-leg ladder extended Bose-Hubbard (BH) model
	\begin{align}
		\label{eq:hamiltonian}
		&\hat{H} =
		-J \sum_{\substack{\Braket{i, j}\\ \alpha\in\{a, b\}}}( \hat{\alpha}_i^\dagger\hat{\alpha}_j + \texttt{H.c.} )
		+\frac{U}{2}\sum_{\substack{i\\ \alpha\in\{a, b\}}} \hat{n}_i^\alpha (\hat{n}_i^\alpha - 1) \nonumber \\
		&\hspace{1cm}  +V \sum_{\substack{i\\ \alpha\in\{a, b\}}} \hat{n}_i^\alpha \hat{n}_{i+1}^\alpha
		+U_{ab}\sum_{i} \hat{n}_i^a\hat{n}_i^b,
	\end{align}
	where, $\hat{\alpha}_i (\hat{\alpha}_i^\dagger)$ is annihilation (creation) operator acting on $i$-th site either ladder $a$ or $b$, and the particle number operator is defined as $\hat{n}^\alpha_i = \hat{\alpha}_i^\dagger\hat{\alpha}_i (\alpha = a,b)$.
	We consider the unit-filling case where $N^\alpha \equiv \sum_i \Braket{\hat{n}_i^\alpha} = L\ (\alpha = a,b)$, with $L$ being the length of the ladder. This is a conserved quantity, as we assume no hopping between the legs, but only between neighboring sites along each leg, with hopping amplitude $J$.
	In each leg, we consider the on-site interaction with strength $U$ and the nearest-neighbor interaction with strength $V$.
	The particles on the same rung but in different legs interact via the inter-leg interaction $U_{ab}$.
	Here, we are particularly interested in the Haldane superfluid phase, which we are targeting using the same system parameters as previous work~\cite{lv_exotic_2016}. Specifically, in both legs we choose $U=1.0$ and $V=0.75$, and tune $J$ in order to reach different phases. 
	A small interaction parameter $U_{ab} = 0.04$ connects the two legs.
	
	The possible phases can be characterized by various quantities.
	\paragraph{Superfluidity.}
	In an infinite homogeneous one-dimensional system with short range interactions and continuous $U(1)$ symmetry, true long-range order and Bose-Einstein condensation breaking the symmetry does not occur due to strong fluctuations. Instead, the SF phase exhibits quasi-long-range order, which reflects in a power-law decay of the SF correlation function. In contrast, the SF correlations over large distances decay exponentially to zero in the insulating phases. The SF correlations in each leg of a 2-leg ladder are defined as
	\begin{gather}
		C^{\alpha}_{SF}(r) =\Braket{\hat{\alpha}_i^\dagger \hat{\alpha}_{i+r}}, 
		\alpha\in \{a, b\}.
		\label{eq:sfcor}
	\end{gather}
	The identification of superfluidity can be challenging since it requires the accurate discrimination of power-law decay from exponential decay which becomes difficult especially close to superfluid to insulating quantum phase transitions.
    %
    Results can strongly depend on the fitting and data analysis approaches.
    Here, we use the value of the correlation function at the largest distance as an indicator of the finite size superfluid "order parameter":
	\begin{gather}
		\tilde{\mathcal{O}}_{SF}^\alpha = C_{SF}^\alpha(r=r_{max}),
		\alpha\in \{a, b\}.
		\label{eq:sfcor_nicco}
	\end{gather}
	
	\paragraph{Density wave phase.} 
	In the conventional one-leg extended BH model at unit-filling, the competition between the on-site and the nearest-neighbor interactions leads to the three different insulating phases: the MI, the DW, and the HI. Strong on-site interaction stabilizes the MI phase, where one particle is localized in each lattice site. However, when the nearest-neighbor interaction is dominant, it is energetically preferable to have a 2-sublattice ground state order where alternating sites have increased and decreased density, respectively,  forming a nearest-neighbor DW order captured by the finite correlation function:
	\begin{gather}
		C^{\alpha}_{DW}(r) =(-1)^r\Braket{\delta\hat{n}_i^\alpha \delta\hat{n}_{i+r}^\alpha}, 
		\alpha\in \{a, b\},
		\label{eq:dwcor}
	\end{gather}
	where $\delta\hat{n}^\alpha_i$ denotes the difference of the particle number at each site from the average particle number in each leg, namely $\delta\hat{n}^\alpha_i = \hat{n}^\alpha_i - 1$. 
	Here, we define the DW order parameter as the value of the correlation function at the largest two-point distance,
	\begin{gather}
		\mathcal{O}^{\alpha}_{DW} = C^\alpha_{DW}(r=r_{\text{max}}), 
		\alpha\in \{a, b\},
		\label{eq:dwop}
	\end{gather}
	
	\paragraph{Haldane insulator.}
	The HI arises when local and non-local interactions are comparable. It is a bulk gapped phase where the DW order described above is diluted (or "hidden") by strings of consecutive sites populated with average density. This phase  resembles the topologically ordered state of spin-1 chains. 
    Other characteristic features of the HI phase are  degenerate edge states, and non-trivial degeneracy in the entanglement spectrum~\cite{li_entanglement_2008,pollmann_entanglement_2010,pollmann_symmetry_2012}.  The most straightforward and well-used quantity for the characterization of the HI is the so-called string order given by
	\begin{align}
		\label{eq:stgcor}
		C^{\alpha}_{STG}(r) &=\Braket{\delta\hat{n}_i^\alpha \exp\left(i\pi\sum_{j\leq k < j+r}\delta\hat{n}_k^\alpha\right) \delta\hat{n}_{i+r}^\alpha},\nonumber\\ 
		&\hspace{4cm}\alpha\in \{a, b\}.
	\end{align}
	The string order parameter can be defined in the same manner as the DW order parameter:
	\begin{gather}
		\mathcal{O}^{\alpha}_{STG} = C^\alpha_{STG}(r=r_{\text{max}}), 
		\alpha\in \{a, b\}.
		\label{eq:stgop}
	\end{gather}
	The spin Haldane phase and the analogous bosonic HI can also be characterized by the degenerate edge mode and the double degeneracy in the lowest entanglement spectrum.
	We primarily compare two tensor network methods for capturing non-local string correlations in the main text.
	The analysis of the edge mode and the entanglement spectrum is presented in the supplemental material.

{\it Results.}
	\begin{figure}[h]
		\centering
		\includegraphics[scale=0.5]{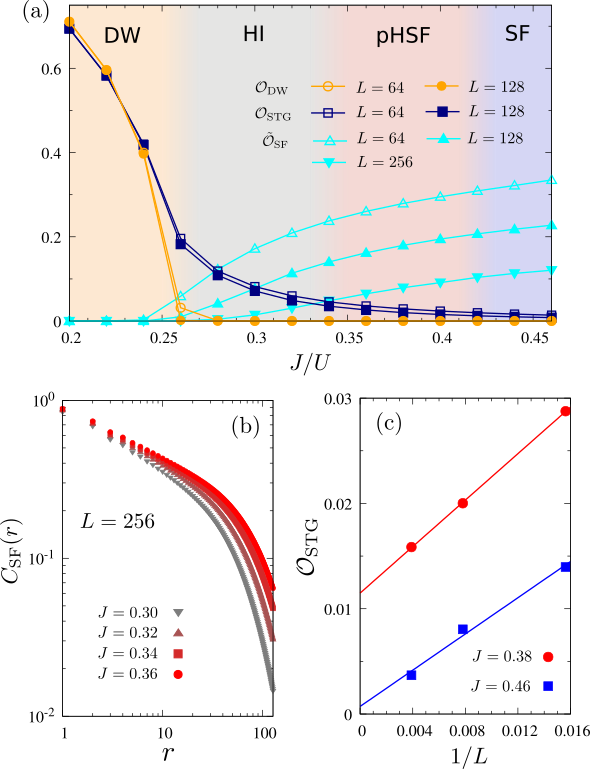}
	\caption{System properties obtained by standard DMRG calculations for the extended Bose-Hubbard two leg ladder with the fixed parameters $U=1.0$, $V=0.75$, and $U_{ab}=0.04$.
			(a) We plot string order parameter (blue lines) and DW order parameter (orange lines) as a function of $J/U$ for $L=64,128$. 
			The SF correlation function at the largest two-point distance is depicted by the cyan- empty triangles ($L=64$), the filled-triangles ($L=128$), and the filled-inverted-triangles ($L=256$).
			By means of these quantities, we differentiate between DW, HI, HSF, and SF phase.
			(b) The SF correlation function for the $L=256$ system size along the HI, the pHSF phases. ($J=0.3\sim 0.36$).
			(c) The system size scaling of the string order parameter $\mathcal{O}_{\text{STG}}$ in the HSF ($J=0.38$, red circles) and the SF ($J=0.46$, blue squares) phases.
		}
		\label{fig:dmrg_orderparams}
	\end{figure}
	\begin{figure*}[t]
		\centering
		\includegraphics[scale=0.52]{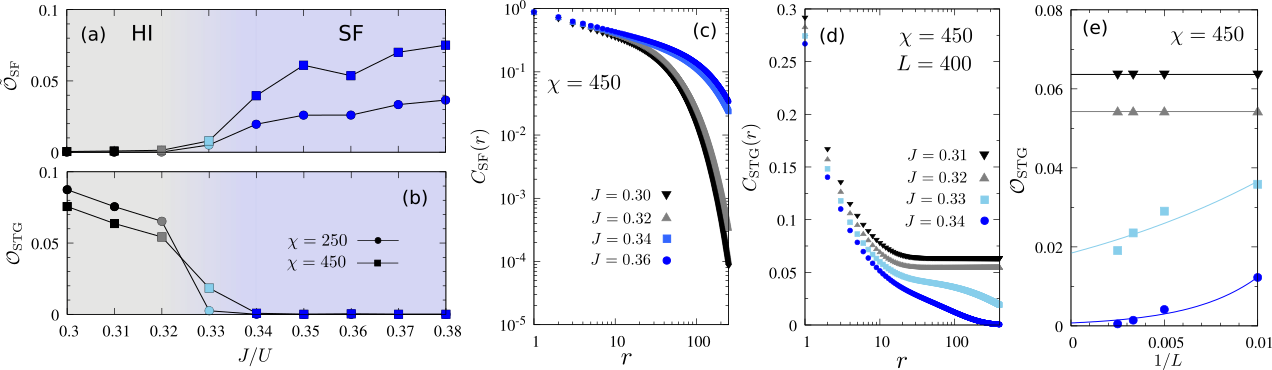}
		\caption{
			 VUMPS results for the SF (a) and the string (b) order parameter with bond dimensions of $\chi=250$ (circles) and $\chi=450$ (squares) across the HI to the pHSF.
			The SF correlation functions near the phase boundary between the HI and the pHSF are shown in panel (c).
			The panels (d) and (e) show the string correlation function $C_{\text{STG}}(r)$ (d) and the scaling of the string order parameter $\mathcal{O}_{\text{STG}}$ (e) near the phase boundary between the HI and the pHSF.
		}
		\label{fig:vumps_stgop}
	\end{figure*}
	We first calculate the ground state properties in  finite systems using the DMRG method.
	In the DMRG calculation, we keep up to $800$ states and require the energy and entropy precisions as $1.0\times 10^{-8}$ and $1.0\times 10^{-5}$, respectively.
	In this paper, we consider open boundary conditions and apply a bias potential, breaking the (edge state's) degeneracy in the DW (HI) phase.
	To this end, we apply additional on-site energy $\epsilon^\alpha_{i}, \alpha\in\{a, b\}$ at the edges $i=1, L$.
	As we consider the repulsive on-site interaction $U_{ab}$ between two legs, it is natural to choose the bias potential to pin the particles alternatively in each leg, namely, $\epsilon^a_1=\epsilon^b_L=0.1, \epsilon^a_L=\epsilon^b_1=2.0$.
	The top panel (a) in Fig.~\ref{fig:dmrg_orderparams} shows how the ground state properties change depending on the hopping for system sizes  $L=64,128,256$.
	The orange circle,  blue rectangle, and cyan triangle lines depict the DW, the string order, and finite size SF order parameters, respectively.
	The system undergoes the transition from the insulating phase to the SF phase depending on the hopping.
	When the hopping is small, the DW and string order parameters are finite, and the system is in the DW phase.
	As the hopping increases, the DW order parameter drops to zero while the string order parameter is finite, showing the transition to the HI phase.
	Even in the SF regime, the string order is still finite, and the system holds the superfluidity and the finite string order parameter simultaneously.
	This phase fulfills the definition of the HSF phase, claimed in Ref.~\onlinecite{lv_exotic_2016}.
	We show the SF correlation function at $L=256$ along the HI, the pHSF phases ($J=0.3\sim 0.36$) in panel (b) in Fig.~\ref{fig:dmrg_orderparams}.
	The tail of the SF correlation function shows the sharper fall in the HI (gray) compared to he pHSF (red) phase, indicating the pHSF regimes exhibit a long-range correlation.
	Panel (c) in Fig.~\ref{fig:dmrg_orderparams} represents the system size scaling of the string order parameter in the pHSF ($J=0.38$) and the SF ($J=0.46$) phases.
	The string order parameter is close to zero at the infinite system size limit in the SF, while it converges to a finite value in the pHSF regime.
	These results are consistent with the QMC calculations (at small finite temperature) with periodic boundary condition~\cite{lv_exotic_2016}. 
	
	Numerical methods for finite systems, such as the DMRG and the QMC, require system size scaling and the artificial choices of boundary conditions for computational accuracy and stability.
	To eliminate these needs, one can consider using periodic boundary conditions (PBC); however, the PBC should usually be avoided in terms of the DMRG due to its high computational costs.
	Another choice is to deal with infinite systems using specialized TN methods, such as the infinite DMRG (iDMRG) and the VUMPS (see supplemental materials for more details).
	The VUMPS globally optimizes and updates the entire system and directly addresses the thermodynamic limit at each iteration without breaking translation invariance, achieving convergence with fewer iterations than the iDMRG, especially for phases with long correlations.
	The superior performance of the VUMPS to capture the long-range correlation over the iDMRG is also shown in the conventional BH model~\cite{kiely_superfluidity_2022}.
	Notably, the VUMPS works efficiently in systems with long-range interactions~\cite{zauner-stauber_variational_2018}, which is essentially important in quasi-1D systems, where effective long-range interactions are introduced by mapping quasi-1D structures into 1D chains.
	For these advantages, we adopt the VUMPS in the following to obtain a more precise phase boundary, especially between the HI and the pHSF, by calculating the SF and the string correlations in infinite systems.

	In Fig.~\ref{fig:vumps_stgop}, we plot the SF and the string order parameters across the HI and the pHSF phases with the maximum bond dimension $\chi = 250, 450$ in panels (a) and (b), respectively.
	The string order parameter in the infinite system is calculated as follows:
	(1) we first calculate the string order parameter in the finite systems, which are composed of copies of the unit cell obtained via the VUMPS algorithm, 
	(2) Then, we conduct the system size scaling to obtain the string order in an infinite system.
	As examples, we show the string correlation function and the system size scaling of the string order parameter in a window of hopping values suggested, by our finite DMRG shown in Fig.\ref{fig:dmrg_orderparams} and previous QMC results \cite{lv_exotic_2016}, to host the phase boundary of the HI and the pHSF in panels (d) and (e) in Fig.~\ref{fig:vumps_stgop}.
	As in the case of the DMRG calculation, we define the superfluidity indicator as values of the SF correlation function Eq.~(\ref{eq:sfcor}) at a certain two-point distance.
	Here, we use $R = 200$, namely,
	\begin{gather}
		\tilde{\mathcal{O}}_{SF}^\alpha = C_{SF}^\alpha(r=R=200),
		\alpha\in \{a, b\}.
		\label{eq:sfcor_nicco_vumps}
	\end{gather}
	We calculate the SF correlation in the same matter as the string order but without the size scaling process (2) due to the nature of quasi-long correlations.
	The SF correlation function across the HI and the pHSF with $\chi = 450$ is shown in panel (c) in Fig.~\ref{fig:vumps_stgop}.
	Interestingly, in contradiction to the previous results, the VUMPS excludes the existence of the pHSF. Indded, the string order rapidly decays to zero  when increasing the hopping in the HI phase (which can also be seen not to host quasi-longrange SF correlations) at $(J/U)_c \approx 0.33$c whereupon the SF correlation begins to increase. This is suggestive of a direct transition between the HI and SF phase, i.e. the previously identified pHSF seems to be an ordinary SF phase. 
    
	We further corroborate our conclusion by analyzing the energy gaps and entanglement spectrum in the supplemental materials which in particular show that the pHSF lacks non-trivial topological features.	
	Studying energy gaps under the open boundary condition and with a bias potential to break the degeneracy in the edge state shows that the pHSF is gapless regardless of the boundary condition.
	While this confirms the superfluid nature of the phase, no fingerprints of topologically non-trivial nature are found in the entanglement spectrum.

	{\it Conclusion.}
	We have investigated the phase transition of an extended BH ladder near the regime where a pHSF of coexisting SF and string orders has been reported \cite{lv_exotic_2016}.
	Using the DMRG and the VUMPS method, we have calculated the string and the SF correlations in finite and infinite systems.
	While the DMRG calculation, even after finite-size scaling, shows an overlap regime of two correlations, the string order sharply vanishes in the VUMPS calculation at the boundary where the SF correlations significantly appear, leaving no room for the pHSF phase. 
	With further investigation of the topological aspects provided in the supplemental material, we conclude that pHSF is, in fact, an ordinary SF phase, contrary to the finite-size DMRG results and Monte Carlo results from Ref.~\onlinecite{lv_exotic_2016}.
	Although our numerical study does not rule out the existence of the HSF phase in other parameter regimes, we speculate that the gapless nature of a superfluid is not compatible with the SPT order of the Haldane phase. 
	Our investigation showcases the advantages of infinite-size tensor network methods for studying topological and gapless phases, as compared to finite-sized scaled DMRG or Monte Carlo methods.
	
	\acknowledgments
    We thank N. Baldelli for advice on the VUMPS calculations.
    
    ICFO-QOT group acknowledges support from:
European Research Council AdG NOQIA;
MCIN/AEI (PGC2018-0910.13039/501100011033, CEX2019-000910-S/10.13039/501100011033, Plan National FIDEUA PID2019-106901GB-I00, Plan National STAMEENA PID2022-139099NB, I00, project funded by MCIN/AEI/10.13039/501100011033 and by the “European Union NextGenerationEU/PRTR" (PRTR-C17.I1), FPI); QUANTERA DYNAMITE PCI2022-132919, QuantERA II Programme co-funded by European Union’s Horizon 2020 program under Grant Agreement No 101017733; Ministry for Digital Transformation and of Civil Service of the Spanish Government through the QUANTUM ENIA project call - Quantum Spain project, and by the European Union through the Recovery, Transformation and Resilience Plan - NextGenerationEU within the framework of the Digital Spain 2026 Agenda;
Fundació Cellex;
Fundació Mir-Puig;
Generalitat de Catalunya (European Social Fund FEDER and CERCA program;
Barcelona Supercomputing Center MareNostrum (FI-2023-3-0024);
Funded by the European Union. Views and opinions expressed are however those of the author(s) only and do not necessarily reflect those of the European Union, European Commission, European Climate, Infrastructure and Environment Executive Agency (CINEA), or any other granting authority. Neither the European Union nor any granting authority can be held responsible for them (HORIZON-CL4-2022-QUANTUM-02-SGA PASQuanS2.1, 101113690, EU Horizon 2020 FET-OPEN OPTOlogic, Grant No 899794, QU-ATTO, 101168628), EU Horizon Europe Program (This project has received funding from the European Union’s Horizon Europe research and innovation program under grant agreement No 101080086 NeQSTGrant Agreement 101080086 — NeQST);
ICFO Internal “QuantumGaudi” project;
    
	TG acknowledges funding by the Department of Education of the Basque Government through the project PIBA\_2023\_1\_0021 (TENINT), and through the IKUR Strategy under the collaboration agreement between the Ikerbasque Foundation and DIPC, by the Agencia Estatal de Investigación (AEI) through Proyectos de Generación de Conocimiento PID2022-142308NA-I00 (EXQUSMI). This work has been produced with the support of a 2023 Leonardo Grant for Researchers in Physics, BBVA Foundation. The BBVA Foundation is not responsible for the opinions, comments, and contents included in the project and/or the results derived therefrom, which are the total and absolute responsibility of the authors.
	YW Ayuda PRE2022-102381 financiada por MCIN/AEI/ 10.13039/501100011033 y por el FSE+. U.B. acknowledges for
the financial support of the IBM Quantum Researcher
Program. R.W.C. acknowledges support from the Polish National Science Centre (NCN) under  Maestro Grant No. DEC- 2019/34/A/ST2/00081.

\newpage

\section{Supplemental Materials}

\subsection{DMRG v.s. VUMPS}	
\subsubsection{Algorithm}
Here, we briefly summarize the algorithm of the two TN methods used in the main text, the DMRG and the VUMPS.
We refer the reader Refs.~\cite{white_density_1992,schollwock_density-matrix_2011} for the DMRG and ~\cite{zauner-stauber_variational_2018,vanderstraeten_tangent-space_2019} for the VUMPS for the further detail.

Consider a general state in an arbitral lattice systems of $L$ sites with $d$ physical dimension,
\begin{align}
	\ket{\Psi} = \sum_{\sigma_1,\sigma_2,\cdots,\sigma_L}c_{\sigma_1,\sigma_2,\cdots,\sigma_L}\ket{\sigma_1,\sigma_2,\cdots,\sigma_L}.
\end{align}
Using the singular value decomposition (SVD), we obtain the matrix product state (MPS) representation in the mixed-canonical form,
\begin{align}
	\ket{\Psi}
	&=
	\sum_{\sigma_1,\sigma_2,\cdots,\sigma_L}
	A_L^{\sigma_1}\cdots A_L^{\sigma_{i-1}}
	A_C^{\sigma_{i}}
	A_R^{\sigma_{i+1}}\cdots A_R^{\sigma_L}
	\ket{\sigma_1,\sigma_2,\cdots,\sigma_L}\nonumber \\
	&=
	\sum_{\sigma_1,\sigma_2,\cdots,\sigma_L}
	A_L^{\sigma_1}\cdots A_L^{\sigma_i}
	C
	A_R^{\sigma_{i+1}}\cdots A_R^{\sigma_L}
	\ket{\sigma_1,\sigma_2,\cdots,\sigma_L}
\end{align}
where $A_C$ is the center site tensor, and $C$ is the bond matrix, a diagonal matrix whose elements are the singular values.
The left/right normalized matrices $A_L$ and $A_R$ satisfy the following relation:
\begin{align}
	\sum_{\sigma_l}A_L^{\sigma_l\dagger}A_L^{\sigma_l} 
	= 
	\sum_{\sigma_l}A_R^{\sigma_l}A_R^{\sigma_l\dagger}
	=
	I.
	\label{eq:relation_normalizedmatrix}
\end{align}
By introducing a constant bond dimension $\chi$ to each tensor $A_{L/R}$, the required parameters can be scaled as $O(Ld\chi^2)$, preventing exponential growth to system sizes.


The DMRG algorithm obtains the MPS representation of states with the minimum energy for the target Hamiltonian $H$ by solving the local eigenvalue problems instead of optimizing the entire state.
To this end, the algorithm optimizes only one or two tensors for each iteration, a process known as one/two-site DMRG.
Due to the larger effective  Hamiltonian, the two-site DMRG is computationally expensive compared to the one-site algorithm, but it is more robust to the local minima and can naturally obtain adequate bond dimensions during the optimization procedure.
Therefore, we mention only the two-site DMRG used throughout this paper.
Consider a tensor created by contracting the neighboring two tensors
\begin{align}
	B^{\sigma_i, \sigma_{i+1}}_{\alpha_{i-1}, \alpha_{i+1}}
	=
	\sum_{\alpha_i}A^{\sigma_i}_{\alpha_{i-1},\alpha_i}
	A^{\sigma_{i+1}}_{\alpha_{i},\alpha_{i+1}},
\end{align}
the effective Hamiltonian acting on the tensor can be written as
\begin{align}
	H_{(\alpha_{i-1},\sigma_i,\sigma_{i+1},\alpha_{i+1})}^{(\alpha_{i-1}^\prime,\sigma_i^\prime,\sigma_{i+1}^\prime,\alpha_{i+1}^\prime)}
	=
	\Braket{\Psi^{\sigma_i^\prime,\sigma_{i+1}^\prime}_{\alpha_{i-1}^\prime, \alpha_{i+1}^\prime}
	|H|
	\Psi^{\sigma_i,\sigma_{i+1}}_{\alpha_{i-1}, \alpha_{i+1}}},
\end{align}
allowing to reduce global optimization to the local two-site eigenvalue problem
\begin{align}
	H_{(\alpha_{i-1},\sigma_i,\sigma_{i+1},\alpha_{i+1})}^{(\alpha_{i-1}^\prime,\sigma_i^\prime,\sigma_{i+1}^\prime,\alpha_{i+1}^\prime)}
	B^{\sigma_i, \sigma_{i+1}}_{\alpha_{i-1}, \alpha_{i+1}}
	=
	E
	B^{\sigma_i, \sigma_{i+1}}_{\alpha_{i-1}, \alpha_{i+1}}.
\end{align}
We obtain the optimal tensor $\tilde{B}^{\sigma_i, \sigma_{i+1}}_{\alpha_{i-1}, \alpha_{i+1}}$ by seeking the ground state eigenvector of the effective Hamiltonian.
The optimized tensor can be restored in the MPS form using the SVD, dynamically adjusting the bond dimension $\chi$.
The DMRG algorithm repeats these procedures sequentially, starting from the left-most sites to the right, and then sweeps backward until it meets the criteria, such as the energy difference between sweeps.

The VUMPS optimizes/updates the tensors in a variational manner by solving the effective eigenvalue problem like sweep-based algorithms, and it follows a very similar algorithm as the DMRG.
At the same time, it updates the entire state globally and directly deals with infinite systems while ensuring translational invariance, leading to its superior performance on systems with long correlations.
The ground state MPS at the thermodynamic limit can be assumed to be translational invariant, and the system is represented by repeating one tensor or a unit-cell containing $N$-tensors. 
The uniform MPS in the mixed-canonical form is given by
\begin{align}
	\ket{\Psi}
	&=
	\sum_{\{\sigma\}}
	\cdots A_L^{\sigma_{i-1}} A_C^{\sigma_i}A_R^{\sigma_{i+1}}\cdots
	\ket{\cdots,\sigma_{i-1},\sigma_{i},\sigma_{i+1},\cdots}\nonumber \\
	&=
	\sum_{\{\sigma\}}
	\cdots A_L^{\sigma_{i}} C A_R^{\sigma_{i+1}}\cdots
	\ket{\cdots,\sigma_{i-1},\sigma_{i},\sigma_{i+1},\cdots}.
\end{align}
The translational invariance of the uniform MPS can be ensured by the relation:
\begin{align}
	A_C = A_L C = C A_R.
\end{align}
To perform the global update while enforcing the translational invariance, we consider the two effective Hamiltonian, acting on $A_C$ and $C$, respectively,
\begin{align}
	&{H_{A_C}}_{\alpha, \sigma_i,\beta}^{\alpha^\prime,\sigma_i^\prime,\beta^\prime}
	=
	\Braket{\Psi^{\sigma_i^\prime}_{\alpha,\beta}
	|H|
	\Psi^{\sigma_i}_{\alpha,\beta}},\nonumber \\
	&{H_{C}}_{\alpha,\beta}^{\alpha^\prime,\beta^\prime}
	=
	\Braket{\Psi_{\alpha^\prime,\beta^\prime}|H|\Psi_{\alpha,\beta}},
\end{align}
leading to the eigenvalue problems for $A_C$ and $C$
\begin{align}
	&{H_{A_C}}_{\alpha, \sigma_i,\beta}^{\alpha^\prime,\sigma_i^\prime,\beta^\prime}A_C
	=
	E_{A_C}A_C,\nonumber\\
	&{H_{C}}_{\alpha,\beta}^{\alpha^\prime,\beta^\prime} C
	=
	E_{C}C.
	\label{eq:vumps_eigenvalueproblem}
\end{align}
Solving both eigenvalue problems gives the optimal tensors $\tilde{A}_C$ and $\tilde{C}$ with the lowest energy of the effective Hamiltonian.
We then update the entire state with the optimal $\tilde{A}_L$ and $\tilde{A}_R$ obtained by minimizing the following errors:
\begin{align}
	\epsilon_L 
	=
	||A_C - A_L C||_2,\ \ 
	\epsilon_R
	=
	||A_C - CA_R||_2.
\end{align}
The bond dimension $\chi$ does not increase during the VUMPS algorithm, as it is based on a one-site DMRG-like scheme.
However, a subspace expansion technique can be used to obtain the desired bond dimension, which is well-explained in~\cite{zauner-stauber_variational_2018}.
The algorithm repeats these optimization processes until it meets the criteria regarding the error given by $\epsilon = \max(\epsilon_L,\epsilon_R)$.

\begin{figure}[htb]
	\centering
	\includegraphics[scale=0.65]{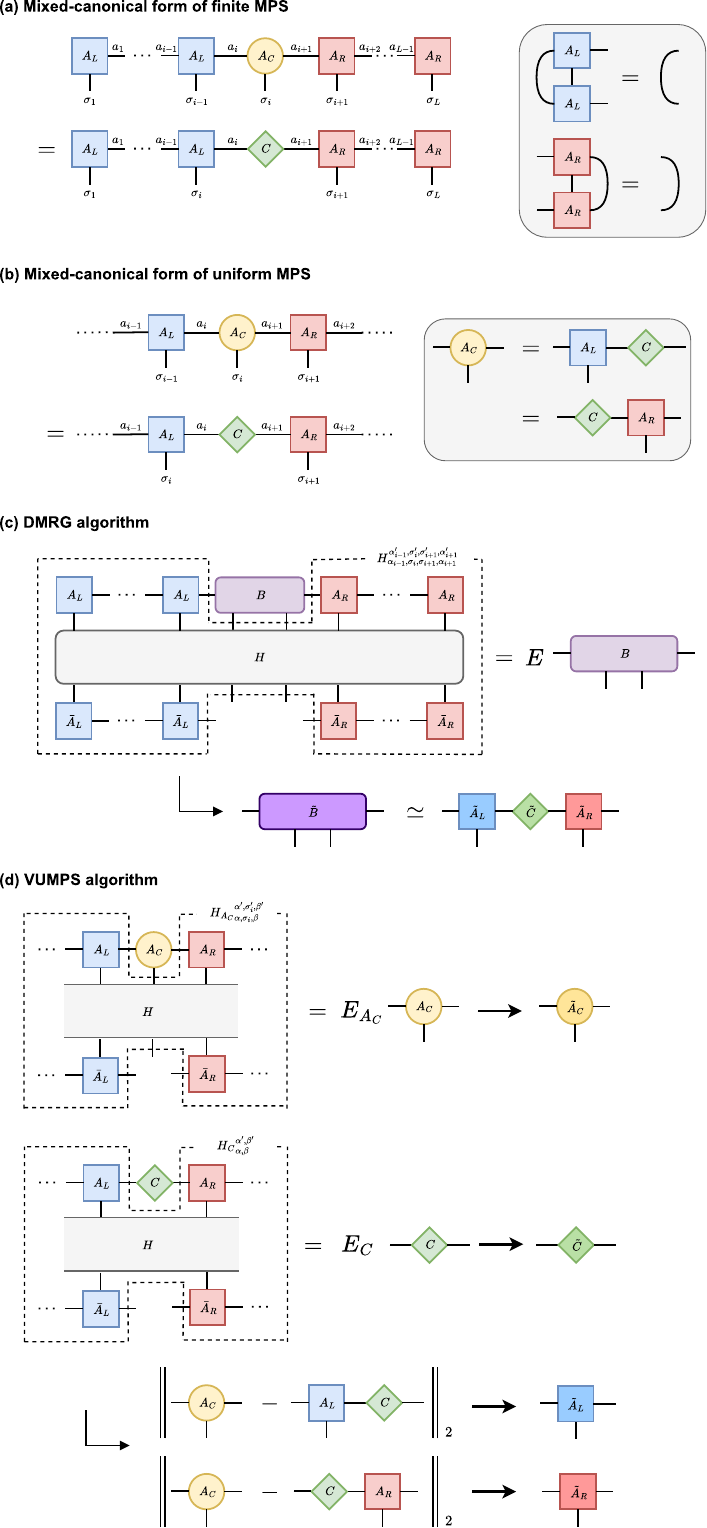}
	\caption{
		Graphical explanation of the two TN methods, the VUMPS and the DMRG, used in our paper.
		Panel (a) and (b) depict the mixed canonical form of the finite MPS (panel (a)) and the uniform MPS (panel (b)).
		The DMRG and the VUMPS algorithms are schematically shown in panel (c) and (d), respectively.
	}
	\label{fig:TN}
\end{figure}

The algorithms mentioned above are schematically shown in Fig.~\ref{fig:TN}.

\subsubsection{Superfluid correlation}
\begin{figure}[H]
	\centering
	\includegraphics[scale=0.6]{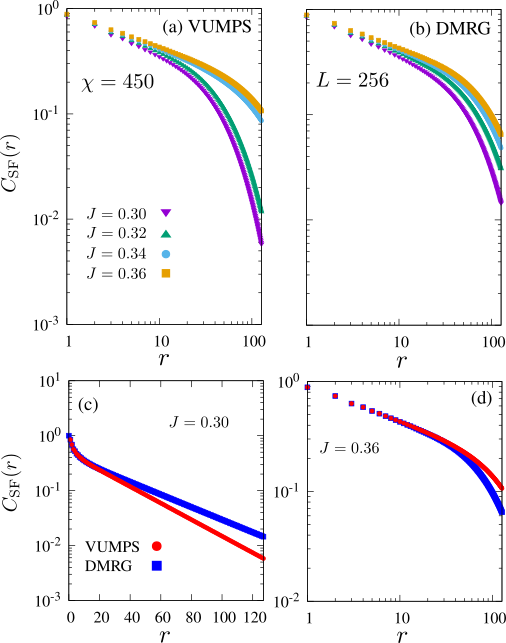}
	\caption{
		The SF correlation function along the HI to the pHSF regimes ($J=0.3 \sim 0.36$) calculated by the VUMPS (panel (a)) and the DMRG (panel (b)).
		In the panel (c) and (d), we show the direct comparison of two TN methods in the HI ($J=0.3$) and the pHSF ($J=0.36$).
		The red circles show the SF correlation function calculated by the VUMPS, while the DMRG calculation is depicted by the blue circles.
	}
	\label{fig:sfcor_dmrgvsvumps}
\end{figure}
As mentioned in the main text, the VUMPS calculation shows that the observables rapidly change depending on phases compared to the DMRG.
This is especially remarkable for long-range correlations.
We plot the SF correlation function calculated by the VUMPS and the DMRG  along the HI to the pHSF regimes ($J/U = 0.3\sim0.36$) in Fig.~\ref{fig:sfcor_dmrgvsvumps}.
Here, we use the same parameters as in the main text.
We allow the VUMPS to keep $\chi=450$ states, and the ground state MPS calculated by the DMRG has up to $\chi=800$ for the $L=256$ system size.
For finite systems, we discard the $L/4$ sites from edges as in the main text, and the SF correlation in Fig.~\ref{fig:sfcor_dmrgvsvumps} is shown up to the largest two-point distance $r = 127$ for both the VUMPS and the DMRG.
We plot the direct comparison of the SF correlator calculated by two TN methods in the HI ($J=0.3$) and the pHSF ($J=0.36$) in panel (c) and panel (d), respectively.
In the pHSF (see panel(d)), the VUMPS shows the algebraic decay of the SF correlation function for the longer two-point distance than the DMRG.
Interestingly, the VUMPS shows a sharper drop of the tail of the SF correlation compared to the DMRG in the HI (see panel (c)).
These differences lead to a sudden change in the behavior of the SF correlation function at the transition point of the HI and the pHSF (between $J=0.32$ and $J=0.34$).

\subsection{Topological properties of the pHSF}
The string order is a good indicator to see if the phase is protected by $\mathbb{Z}_2\times\mathbb{Z}_2$ symmetry; however, this quantity is not enough to capture the presence of the other symmetry and the corresponding topological properties.
We provide additional data to characterize the topological phases other than the string order shown in the main text.
We calculate the energy gaps in the finite systems with different boundary conditions to show the characteristic properties of the SPT phase, such as the gapped bulk and the gapless edge modes, do not appear in the pHSF phase.
Further investigation examines topological properties from the entanglement point of view and shows the absence of topological 'fingerprints' throughout the pHSF and the SF phases.

\subsubsection{Edge mode}
\begin{figure}[H]
	\centering
	\includegraphics[scale=0.3]{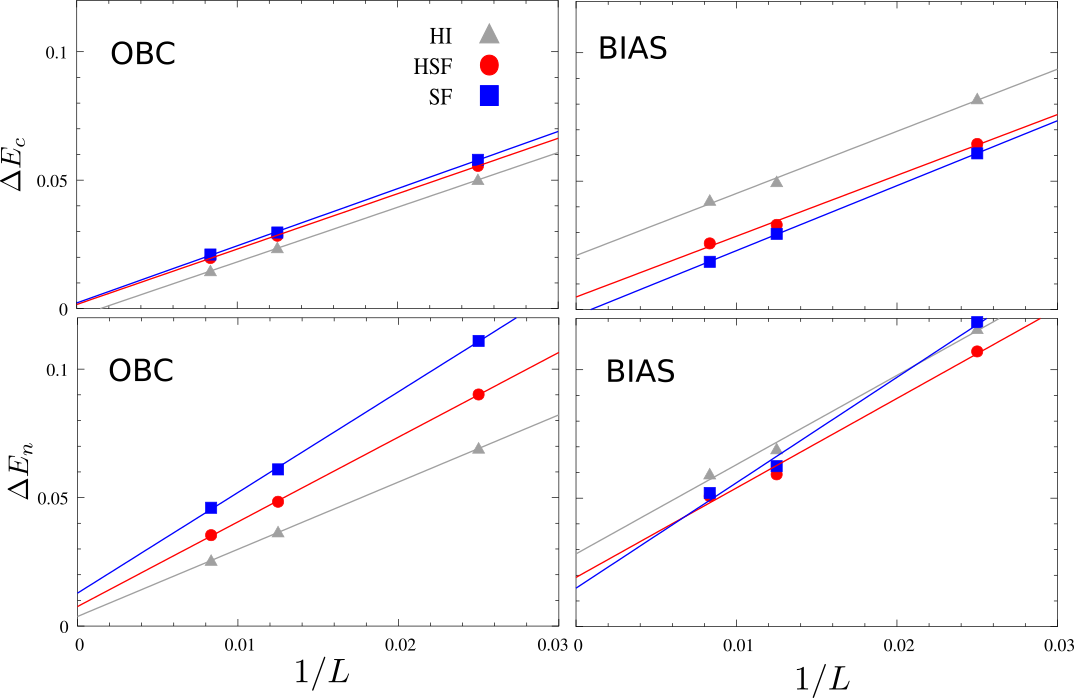}
	\caption{
		The system size scaling of the charge (upper panels) and neutral (lower panels) energy gaps with the open boundary condition (left panels) and with the bias potential applied (right panels) in the HI (gray triangles), the HSF (red circles), and the SF (blue squares) phases.
	}
	\label{fig:energygap}
\end{figure}
\begin{figure*}[t]
	\centering
	\includegraphics[scale=0.40]{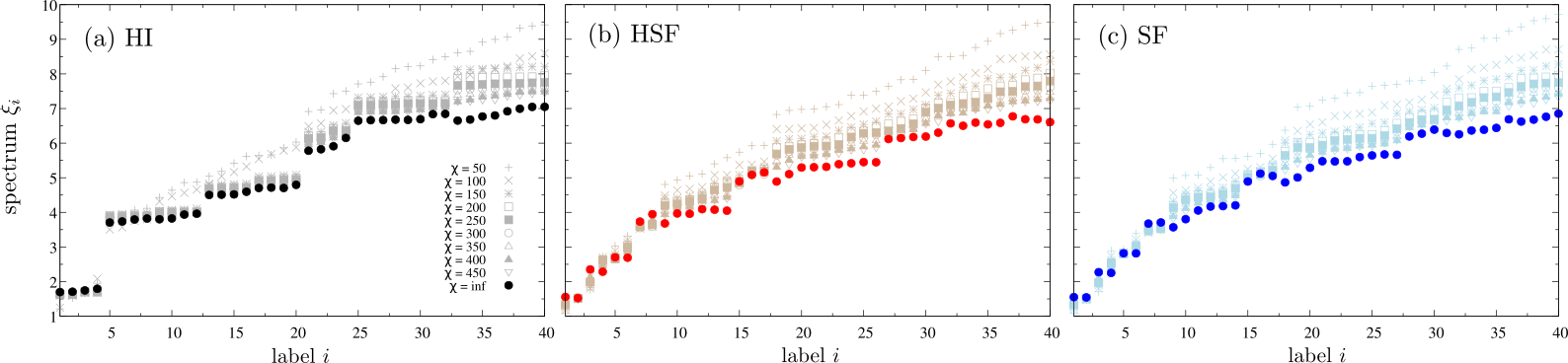}
	\caption{
		The lowest entanglement spectrum in the infinite system in the HI (panel (a)), the pHSF (panel (b)), and the SF (panel (c)).
		The spectrum at the infinite bond dimension (filled circles) is obtained by the scaling with $\chi = 50\sim 450$.
	}
	\label{fig:spectrum}
\end{figure*}
The presence of the edge mode can be studied in several ways.
We calculate the energy gaps with or without the bias potential and see if the differences arise due to the presence of the edge states.
The same bias potential as in the main text is used to break the degeneracy of the edge state.
We consider the charge and the neutral energy gaps $\Delta E_c, \Delta E_n$, which are defined as follows respectively:
\begin{align}
	\Delta E_c &= E^{(0)}(N_a+1, N_b) + E^{(0)}(N_a-1,N_b)\nonumber\\
	&\hspace{3cm}- 2 E^{(0)}(N_a,N_b),\\
	\Delta E_n &=  E^{(1)}(N_a,N_b) - E^{(0)} (N_a,N_b),
\end{align}
where, $E^{(0)}(N_a,N_b)$ represents the ground energy with $N_a$ and $N_b$ particles in respective ladder $a$ and $b$, and $ E^{(1)}(N_a,N_b)$ denotes the first excitation energy.
We note here that there are several ways to define the charge energy gap in the two-leg ladder model.
However, all the definitions give only a constant factor difference, as long as both legs are identical and the interaction between them is sufficiently small that the total energy can be written in terms of the sum of the individual contributions from each leg.

We show the charge and the neutral energy gaps in Fig.~\ref{fig:energygap}.
We use the same parameters as in the main text, but $J/U = 0.30$, $J/U = 0.38$, and $J/U = 0.50$ for the HI, the pHSF, and the SF, respectively.
The left panels show the charge (upper) and the neutral (lower) energy gaps calculated under the open boundary condition in the HI (gray triangles), the pHSF (red circles), and the SF phases (blue squares).
The right panels depict the results with the bias potential applied.
These two different boundary conditions reveal the topological properties specific to the HI phase: The phase is gapless with the open boundary condition. However, the gapped bulk appears when the degeneracy in the edge state is broken due to the bias potential.
However, the pHSF shows the same behavior as the SF; the phase is gapless regardless of the bias potential.
This observation of a gapless bulk in the pHSF re-affirms the presence of superfluidity in the pHSF regime.

\subsubsection{Entanglement spectrum}

The entanglement spectrum gives a comprehensive point of view to the topological properties in the system. Specifically, the entanglement spectrum shows a double degeneracy in the lowest spectrum in the spin Haldane phase and the bosonic HI.
Consider dividing the system with $L$ lattice sites into two sub systems $A$ and $B$ with size of $l$ and $L$, we obtain the reduced density matrix $\rho_A = \text{Tr}_{B}\ket{\Psi}\bra{\Psi}$.
The entanglement spectrum is given as $\xi_i = -2\ln\lambda_i$, where $\lambda_i$ is the eigenvalue of the reduced density matrix $\rho_A$.
The entanglement spectrum can be easily computed using the relation $\lambda_i = S_i^2$, where $S_i^2$ is the singular value we can easily access in terms of matrix product state.

We calculate the entanglement spectrum using the VUMPS method in Fig.~\ref{fig:spectrum}.
We plot the lowest part of the entanglement spectrum $\xi_i, (i = 1\sim 40)$ in the HI regime (Fig.~\ref{fig:spectrum}(a)), the pHSF regime (Fig.~\ref{fig:spectrum}(b)), and the SF regime (Fig.~\ref{fig:spectrum}(c)).
The used parameters are the same as in the main text, but $J/U = 0.30$, $J/U = 0.38$, and $J/U = 0.46$ for the HI, the pHSF, and the SF, respectively.
The filled circles represent the spectrum at the infinite bond dimension $\chi\rightarrow\infty$ obtained by scaling the data with the different bond dimension $\chi = 50\sim 450$.
Here, we note that all the phases show the double-degeneracy in the lowest spectrum due to the two-leg ladder structure.
In the HI phase, we can see the double- or four-degeneracy, and these fingerprints are well consistent with the HI in the one-leg ladder model~\cite{ejima_spectral_2014}, and with the Haldane phase in spin systems~\cite{pollmann_entanglement_2010}.
However, the pHSF shows similar fingerprints as the SF, and no topological properties are seen in the entanglement spectrum. This observation is in line with vanishing string order parameter, observed from the VUMPS calculation in the pHSF regime.

\bibliography{bibliography.bib}

\end{document}